\documentclass{aa}  
\usepackage{graphicx}
\usepackage{txfonts}
\def\xcma{$\xi^1$\,CMa}
\def\tcar{$\theta$~Car}
\def\tsco{$\tau$\,Sco}

\begin{document}
   \title{New insights into the nature of the peculiar star
$\theta$~Carinae\thanks{Based
on observations collected with the CORALIE spectrograph attached to the Euler Telescope 
of the Geneva Observatory situated at La Silla  (Chile), at the European Southern 
Observatory, Paranal, Chile
(ESO programmes 67.D-0239(A), 072.D-0377(A), 078.D-0080(A) and 278.D-5056(A)),
and at the Complejo Astron\'omico El Leoncito (CASLEO), Argentina.
}
}

   \author{
S. Hubrig\inst{1}
\and
M. Briquet\inst{2}\thanks{Postdoctoral Fellow of the Fund for Scientific Research, Flanders}
\and
T. Morel\inst{2,3}
\and
M. Sch\"oller\inst{1}
\and
J. F. Gonz\'alez\inst{4}
\and
P. De Cat\inst{5}
}

   \offprints{S. Hubrig}

   \institute{
European Southern Observatory, Casilla 19001, Santiago 19, Chile
\and
Instituut voor Sterrenkunde, Katholieke Universiteit Leuven, Celestijnenlaan 200 D, B-3001 Leuven, Belgium
\and
Institut d'Astrophysique et de G\'eophysique, Universit\'e de Li\`ege, All\'ee du 6 Ao\^ut, B\^at. B5c, 4000 Li\`ege, Belgium
\and   
Complejo Astron\'omico El Leoncito, Casilla 467, 5400 San Juan, Argentina
\and
Koninklijke Sterrenwacht van Belgi\"e, Ringlaan 3, B-1180 Brussel, Belgium
}

   \date{Received Enero 01, 2008; accepted Enero 31, 2008}

 
  \abstract
   { $\theta$~Carinae belongs to a group of peculiar early-type stars (OBN) with enhanced nitrogen and carbon deficiency.
It is also known as a binary system, but it is not clear yet whether the chemical anomalies can be 
explained by mass transfer between the two components.
On the basis of the previously reported spectral 
variability of a few metal lines it may be expected that \tcar{} possesses a weak magnetic field.  }
   {A study of the physical nature of this hot massive binary which is furthermore a well-known 
blue straggler lying $\sim$2\,mag above the turnoff of the young open cluster IC~2602 is important to 
understand the origin of its strong chemical anomalies.}
   { We acquired high resolution spectroscopic and low resolution spectropolarimetric observations to 
achieve the following goals:
a) to improve the orbital parameters to allow a more 
in-depth discussion on the possibility of mass transfer in the binary system,
b) to carry out 
a non-local thermodynamic equilibrium (NLTE) abundance analysis, and
c) to search for the presence 
of a magnetic field.  }
   {
The study of the radial velocities using CORALIE spectra 
allowed us to significantly improve the orbital parameters.
A comparative NLTE abundance analysis was undertaken for \tcar{} and two other  
early B-type stars with recently detected magnetic fields, \tsco{} and \xcma{}. 
The analysis revealed significantly different abundance patterns:
a one-order-of-magnitude nitrogen overabundance and carbon depletion was found in \tcar{}, while the oxygen 
abundance is roughly solar. For the stars \xcma{} and \tsco{} the carbon abundance 
is solar and, while an N excess is also detected, it is of much smaller amplitude (0.4--0.6\,dex). Such an 
N overabundance is typical of the values already found for other slowly-rotating (magnetic) B-type dwarfs.
For \tcar{}, we attribute instead the chemical peculiarities to a past episode of mass transfer
between the two binary components.
The results of the search for a magnetic field using FORS\,1 at the VLT consisting of 26 measurements 
over a time span of $\sim$1.2\,h are rather inconclusive: only few measurements have a significance level of 3$\sigma$.
Although we detect a periodicity of the order of $\sim$8.8\,min in the 
dataset involving  the measurements on all hydrogen Balmer lines with the exception of 
the H$\alpha$ and H$\beta$ lines, 
these results have to be confirmed by additional time-resolved magnetic field observations.
   }
  {}

   \maketitle

%

\section{Introduction}
\label{sect:intro}

It has long been assumed that massive stars do not have magnetic fields, as they lack 
the convective outer mantle prevalent in lower mass stars.
However, indirect evidence
is supporting that magnetic fields are indeed present in massive O and 
early B-type stars 
(e.g., Henrichs et al.\ \cite{henrichs}; Rauw et al.\ \cite{rauw};
Cohen et al.\ \cite{cohen}; Gagn{\'e} et al.\ \cite{gagne}).
Yet, only very few direct magnetic field detections 
have been reported in O-type stars so far (Donati et al.\ \cite{donati06a}; Wade et al.\ \cite{wade06}; Hubrig et al.\ \cite{hubrig07}).
Among the hottest B-type stars, a magnetic field has been discovered in the B0.2V star \tsco{}
(Donati et al.\ \cite{donati06b}) and in the B0.7IV star \xcma{}, which is
one of the hottest $\beta$\,Cephei stars, with a
rather large longitudinal magnetic field of up to 300\,G (Hubrig et al.\ \cite{hubrig06}).

Walborn (\cite{walborn06}) listed the hot B0.2V star $\theta$~Carinae (HD\,93030, HIP\,52419, HR\,4199; $m_V$ = 2.78)  among a 
few other massive stars with unexplained spectral peculiarities or variations for which a magnetic field could be expected.
It has a peculiar, variable spectrum in both optical and UV.
The spectral peculiarities in the blue-violet are an enhancement of nitrogen and 
deficiency of carbon, but also definite line-intensity variations as well as other line
asymmetries (Walborn \cite{walborn76,walborn79}).

The non-detection of a magnetic field was reported by Borra \& Landstreet (\cite{borra_land}), who used
a photoelectric Balmer-line magnetograph to measure circular polarization in 
the wings of the H$\beta$ line. Although \tcar{} is a very bright target, easily
observable with spectropolarimeters, no other magnetic field measurements have been reported so far 
in the literature, mainly due to the unavailability of instruments  equipped with polarization analyzing 
optics on telescopes located in the southern hemisphere.

\tcar{} is an SB1 system with one of the shortest orbital periods
known among massive stars ($P$ = 2.2\,d; Lloyd et al.\ \cite{lloyd}).
A discussion of the possibility of 
mass transfer in the binary system, which would be a natural explanation for the remarkable 
spectral peculiarities and for the singular location of this object in the H-R diagram of 
the 30\,Myr old open cluster IC\,2602, was presented by Walborn (\cite{walborn76}). 
However, the spectral type of the companion remains unknown. Previously determined orbital 
elements were rather uncertain 
(Lloyd et al.\ \cite{lloyd}) and other periods of $\sim$1\,day and $\sim$25~days have been suggested
due to a systematic difference in the radial velocities 
from different observers.

Randich et al.\ (\cite{randich}) reported in their ROSAT/PSPC study of the  
cluster IC\,2602 that  \tcar{}
is the brightest X-ray object
among the studied cluster
members, with a soft X-ray luminosity amounting to up to $\log L_{\rm X}$ = 30.99\,ergs\,s$^{-1}$.
{
Recently, Naz\'e \& Rauw (submitted), using XMM-{\it Newton} observations, showed that 
the X-ray flux  of \tcar{} is slightly 
lower than the flux typically observed in O and early B-type stars and confirmed the unusual softness 
of the X-ray emission.
Further, they noted that X-ray lines appear narrow and unshifted, reminiscent of those of $\beta$\,Cru
and the magnetic star \tsco{} (Donati et al.\ \cite{donati06b}).
}

Below, we present the results of our new spectroscopic and spectropolarimetric study of 
this peculiar massive star and discuss possible origins of its anomalies.

\section{Orbital elements of the SB1 system}
\label{sect:binarity}
\begin{table}
\centering
\caption{
Individual radial velocity measurements. 
Observations  between MJD\,54303 and MJD\,54313 were taken with CORALIE,
the CASLEO spectra between MJD\,54520 and MJD\,54581, while  MJD\,54141 and 52027 
correspond to FEROS spectra.
}
\label{tab:rv}
\begin{tabular}{ccc|ccc}
\hline
\hline
MJD & Phase & RV & MJD & Phase & RV \\
 & & km/s & & & km/s \\ \hline

52027.0699 & 0.0939 &  06.00 & 54311.0013 & 0.8879 &  35.49 \\ 
54141.2248 & 0.8176 &  37.70 & 54311.0051 & 0.8896 &  35.32 \\ 
54303.9597 & 0.6913 &  36.03 & 54311.0090 & 0.8914 &  34.84 \\ 
54303.9902 & 0.7052 &  36.51 & 54311.0129 & 0.8931 &  34.66 \\ 
54304.0025 & 0.7108 &  36.95 & 54311.0168 & 0.8949 &  34.54 \\ 
54305.9760 & 0.6066 &  29.34 & 54311.0206 & 0.8966 &  34.60 \\ 
54305.9832 & 0.6099 &  29.43 & 54311.0245 & 0.8984 &  34.33 \\ 
54305.9967 & 0.6160 &  30.04 & 54311.0283 & 0.9001 &  34.17 \\ 
54306.0049 & 0.6198 &  30.42 & 54311.0323 & 0.9020 &  33.92 \\ 
54306.0165 & 0.6250 &  30.73 & 54311.0366 & 0.9039 &  33.93 \\ 
54306.0250 & 0.6289 &  31.28 & 54311.0409 & 0.9059 &  33.92 \\ 
54306.0342 & 0.6331 &  31.29 & 54311.0451 & 0.9078 &  33.35 \\ 
54306.0400 & 0.6357 &  31.65 & 54311.0493 & 0.9097 &  33.48 \\ 
54306.0463 & 0.6386 &  31.69 & 54311.0535 & 0.9116 &  33.28 \\ 
54307.9612 & 0.5078 &  20.10 & 54311.9615 & 0.3238 &  05.08 \\ 
54307.9652 & 0.5096 &  20.24 & 54311.9701 & 0.3277 &  05.38 \\ 
54307.9694 & 0.5115 &  20.31 & 54311.9737 & 0.3293 &  05.47 \\ 
54307.9801 & 0.5164 &  20.89 & 54311.9777 & 0.3311 &  05.63 \\ 
54307.9960 & 0.5236 &  21.46 & 54311.9843 & 0.3341 &  05.83 \\ 
54308.0073 & 0.5288 &  21.91 & 54311.9921 & 0.3377 &  05.87 \\ 
54308.9867 & 0.9734 &  24.61 & 54311.9958 & 0.3393 &  06.05 \\ 
54308.9907 & 0.9752 &  24.35 & 54312.0112 & 0.3463 &  06.59 \\ 
54308.9949 & 0.9771 &  24.24 & 54312.0199 & 0.3503 &  06.93 \\ 
54308.9991 & 0.9790 &  23.88 & 54312.0299 & 0.3548 &  07.21 \\ 
54309.0029 & 0.9807 &  23.46 & 54312.0335 & 0.3564 &  07.41 \\ 
54309.0069 & 0.9825 &  23.32 & 54312.0390 & 0.3589 &  07.86 \\ 
54309.0109 & 0.9843 &  22.71 & 54312.0425 & 0.3605 &  07.92 \\ 
54309.0147 & 0.9861 &  22.67 & 54312.0471 & 0.3626 &  07.90 \\ 
54309.0186 & 0.9878 &  22.43 & 54312.0506 & 0.3642 &  08.20 \\ 
54309.0223 & 0.9895 &  22.30 & 54520.2392 & 0.8717 &  37.63 \\
54309.0260 & 0.9912 &  21.83 & 54521.2403 & 0.3262 &  05.73 \\
54309.0297 & 0.9929 &  21.72 & 54522.2258 & 0.7736 &  36.88 \\
54310.9623 & 0.8702 &  36.78 & 54544.2359 & 0.7651 &  38.05 \\
54310.9662 & 0.8719 &  36.79 & 54545.2886 & 0.2429 &  02.74 \\
54310.9700 & 0.8737 &  36.79 & 54546.2425 & 0.6759 &  36.26 \\
54310.9745 & 0.8757 &  36.39 & 54576.1355 & 0.2459 &  01.58 \\ 
54310.9783 & 0.8774 &  36.18 & 54577.1151 & 0.6906 &  34.80 \\ 
54310.9821 & 0.8792 &  36.25 & 54577.1842 & 0.7220 &  37.22 \\ 
54310.9860 & 0.8809 &  36.03 & 54578.0030 & 0.0937 &  07.68 \\ 
54310.9899 & 0.8827 &  35.83 & 54578.1381 & 0.1550 &  03.42 \\ 
54310.9937 & 0.8844 &  35.91 & 54579.1201 & 0.6008 &  26.86 \\ 
54310.9974 & 0.8861 &  35.29 & 54580.1620 & 0.0738 &  11.24 \\ 
\hline
\end{tabular}
\end{table}

\begin{figure}
\centering
\includegraphics[width=0.45\textwidth]{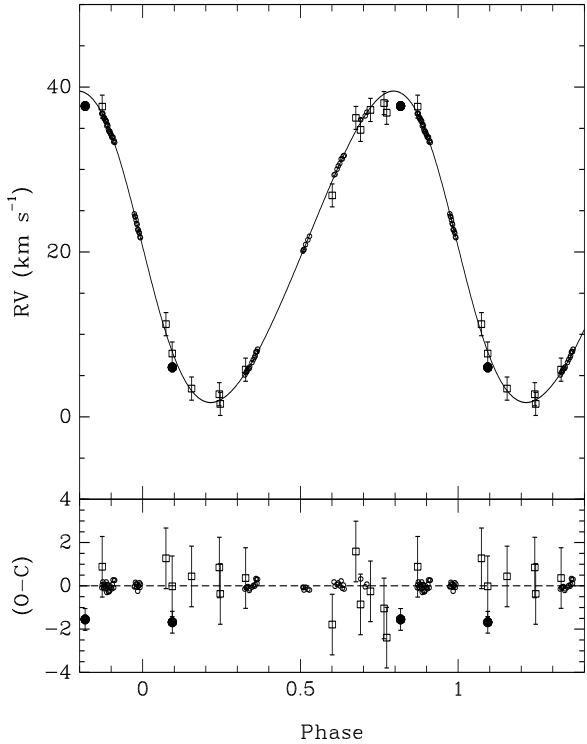}
\caption{Radial velocity curve of \tcar{}.
Small open circles are observations from CORALIE, open squares from CASLEO,
and the filled circles are from FEROS. 
The solid line corresponds to the adopted orbital solution.
Residuals (O-C) are shown in the lower panel. 
}
\label{fig:orb}
\end{figure}

\begin{figure}
\centering
\includegraphics[width=0.4\textwidth,angle=270]{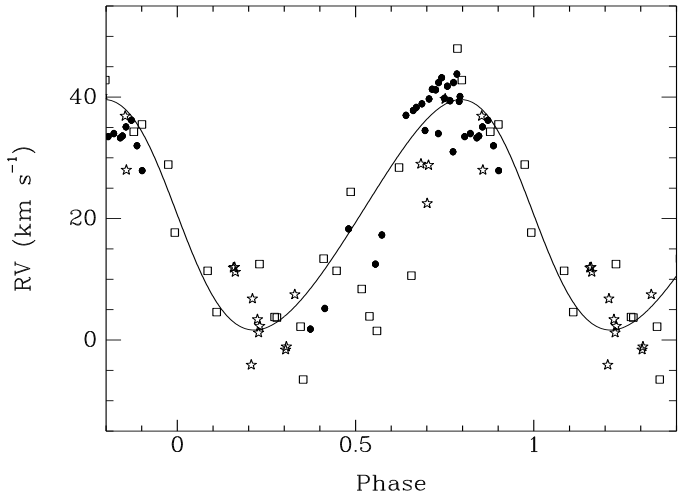}
\caption{Comparison of old radial velocity measurements with our calculated RV curve. 
Open squares: Walborn (\cite{walborn79}) shifted by $-$10\,km\,s$^{-1}$,
open stars:  Walker \& Hill (\cite{walker85}) shifted by $+$10\,km\,s$^{-1}$,
filled circles: Lloyd et al.\ (\cite{lloyd}) shifted by $+$15\,km\,s$^{-1}$.
}
\label{fig:orbold}
\end{figure}

\begin{table}
\centering
\caption{Orbital elements for \tcar{}.
$P$ is the orbital period, 
$v_\gamma$ is the system velocity,
$K_1$ the semi-amplitude of the radial velocity curve of component 1,
$e$ is the eccentricity,
$\omega_1$ the periastron longitude,
$a_1$ the semi-major axis,
$i$ the orbital inclination,
$t({\rm conj})$ the time of the primary conjunction, and
$t({\rm per})$ the time of the periastron passage.
} 
\label{tab:orbit}
\begin{tabular}{lr@{}c@{}l}
\hline
\hline
$P$           & 2.20288    & \,$\pm$\, & 0.00001\,d \\
$v_\gamma$    & 20.18       & $\pm$ & 0.04\,km\,s$^{-1}$ \\
$K_1$         & 18.93       & $\pm$ & 0.05\,km\,s$^{-1}$ \\
$e$           & 0.129      & $\pm$ & 0.002 \\
$\omega_1$    & 81.8         & $\pm$ & 1.7\,$^\circ$ \\
$a_1$ sin $i$ & 0.00395    & $\pm$ & 0.00014\,AU \\
$t({\rm conj})$     & MJD\,54302.4367& $\pm$ & 0.0012 \\
$t({\rm per})$     & MJD\,54302.3984& $\pm$ & 0.0092 \\
\hline
\end{tabular}
\end{table}

The radial velocity (RV) of \tcar{} has been discovered to be variable almost a century ago 
(Wilson \& Sanford \cite{wilson}). Although binarity is considered as the most plausible 
explanation for the observed short-period variations, the orbit was not well known prior to our study.
The most recent orbit determination is given by Lloyd et al.\ (\cite{lloyd}), who found a circular 
orbit with a period of 2.203020\,d and an amplitude of 14.4\,km\,s$^{-1}$. It is based on all 
spectroscopic data that were available at that time. They found evidence for an additional 
period of about 25~days. However, they noted that both periods should be checked with supplementary data. 
To improve the orbit, 69~high-resolution ($R$ = 47\,000) high signal-to-noise spectra 
(S/N $\sim$200 at 4500\,\AA{}) 
were taken  with the CORALIE spectrograph attached 
to the 1.2\,m Euler telescope (La~Silla, Chile)
from 2007 July 22 to 30.
Exposure times were in the range 100--200 seconds.
{Thirteen} additional spectra were taken between February and May 2008
at the CASLEO with the 2.1\,m telescope and the REOSC spectrograph ($R$ = 13\,000)
to improve the period determination.
We also included in the radial velocity analysis a FEROS spectrum ($R$ = 48\,000)
obtained in February 2007 which was used for the chemical abundance analysis 
(see Sec.~\ref{sect:abundances}) and 
a FEROS spectrum obtained for the program 67.D-0239(A) in April 2001.

Radial velocities were measured by cross-correlations with a high-S/N spectrum of
a star of similar spectral type (HD\,37042, B1V).
The RV error of this template is about 0.5\,km\,s$^{-1}$, which we consider as the formal error
of the zero point.
The residuals of the fitted orbit indicate that the probable error in relative radial
velocities is smaller (see below).
{During the orbit fitting we noted that CASLEO radial velocities show a systematic 
difference of
about 0.9\,km\,s$^{-1}$ relative to CORALIE observations.
To correct possible
differences in the instrumental zero point, we measured the interstellar \ion{Na}{i}  doublet lines
at $\lambda\lambda$5890, 5896. For these interstellar lines we 
obtained a mean radial velocity of 8.39$\pm$0.01\,km\,s$^{-1}$ for CORALIE spectra and 
9.42$\pm$0.37\,km\,s$^{-1}$ for CASLEO
spectra. Consequently, we applied a correction of $-$1.03\,km\,s$^{-1}$ to the measured 
CASLEO velocities.} 

A Keplerian orbit was fitted by least-squares to the  measured RVs
listed in Table~\ref{tab:rv}.
Different weights have been assigned to the CORALIE and CASLEO observations due to their
unequal probable errors:
0.15\,km\,s$^{-1}$ and 1.2\,km\,s$^{-1}$, respectively.
These values were estimated from the residuals of the fitting itself.
{We found an orbital period of 2.20292$\pm$0.00005\,d.
Once all orbital parameters were determined, we included in the analysis 
radial velocities previously published by Walborn (\cite{walborn79}), 
Walker \& Hill (\cite{walker85}), and Lloyd et al.\ (\cite{lloyd}). 
These old radial velocities can be fitted well with our radial velocity curve modifying
slightly the period to $P$ = 2.20288\,d. 
Finally, we kept the period fixed and re-fitted the remaining orbital parameters with our
radial velocities.}
The orbital solution is plotted in Fig.~\ref{fig:orb} and
the orbital elements are presented in Table~\ref{tab:orbit}.
All our observations, which span over one year,
are consistent, within the errors, with a single orbit. The error in the center of mass velocity is 
just the formal error of the fit, but the absolute error is larger.
Besides, different lines in \tcar{} could possibly have slightly different velocities due to 
low-level spectral variability detected by Walborn (\cite{walborn76}). Thus it is not easy to
decide which is the correct center-of-mass velocity.

The determined orbital period is in good agreement with the period found by Lloyd et al.\ (\cite{lloyd}).
However, we find a significant eccentricity ($e$ = 0.13) and a higher
amplitude ($K_1$ = 18.9\,km\,s$^{-1}$).
Furthermore, we do not find any definite evidence for the presence of 
secular changes in the orbital parameters or additional velocity variations,
to a level of about 1\,km\,s$^{-1}$.

{In Figure~\ref{fig:orbold} we present the older velocity measurements from the literature 
compared with our adopted orbit.
Arbitrary velocity shifts have been applied to all three
data samples as indicated in the figure caption. 
It is possible that the origin of these systematic differences is related to the
use of spectral lines of different chemical elements and in different spectral regions for the measurements 
of radial velocities, although
a long period variability cannot be completely discarded.}

Due to the considerable number of CORALIE observations and the weights assigned, 
the orbital parameters are mainly determined by the CORALIE data and are representative of the
orbit at the epoch of CORALIE observations, while the FEROS and CASLEO spectra 
mainly contribute to a more precise determination of the orbital period.
The sampling of our CORALIE data, however, is not adequate to allow us to detect a period as long as 25~days,
which was mentioned by Lloyd et al.\ (\cite{lloyd}).

{The secondary component is not visible in our CORALIE spectra. 
In order to search for the secondary spectrum we proceeded in following way:
First, each observed spectrum was shifted in wavelength by the amount 
corresponding to the primary velocity to allow the calculation of
the mean spectrum of the primary component. This mean spectrum was subtracted from each 
observed spectrum.
The residual spectra are expected to be low-S/N  spectra of the secondary star. 
To combine them it is necessary to know the secondary radial velocities, which are not a priori known.
On the other hand, the radial velocity curve of the secondary can be calculated from the primary curve 
if a value for the mass-ratio is adopted.
For a grid of mass-ratios in the range 0.03--0.60 we calculated the corresponding velocity of the
secondary star for the time of each observation, and applied corresponding shifts to the residual 
spectra. The average secondary spectrum was calculated for each assumed mass-ratio. 
Finally we cross-correlated these spectra against templates in the range of spectral types A7--F9, 
which correspond to the temperatures expected for main-sequence stars of about 1--2 solar
masses (see Sec.\,5).
For all considered mass-ratio values we find no evidence for the presence of a cross-correlation-peak 
corresponding to the secondary star.
From tests with synthetic spectra with noise we estimate that a slowly rotating F-type star can
be detected  by the cross correlation even for a S/N-ratio as low as 1.3. 
Since the mean spectrum is expected to have a S/N ratio of about 1400 with respect to the 
combined continuum, we conclude that, if the secondary star is an F-type star with moderate rotation, 
then the secondary spectrum contributes less than 0.1\% to the total flux.}

\section{Abundance analysis and spectral variability}
\label{sect:abundances}

NLTE abundances of helium and several metals were 
calculated using the latest versions of the line formation codes DETAIL/SURFACE and plane-parallel, fully 
line-blanketed model atmospheres (Kurucz \cite{kurucz}).
Note that ignoring the low luminosity companion in \tcar{} has no bearing on our 
abundance analysis. For comparison purposes, we also derived the abundances of the two magnetic
early B-type dwarfs \xcma{} and \tsco{}, following exactly 
the same procedures. Atmospheric models with a solar helium abundance were adopted, in accordance with the 
values found for all targets (see below). Our analysis of \tcar{} is based on a FEROS spectrum obtained on 
2007 February 10 with the ESO/MPI 2.2\,m telescope.
This spectrum has a resolution similar to the CORALIE spectra used for the orbit determination in Sect.~\ref{sect:binarity}, but
has a higher signal-to-noise ratio than the individual CORALIE spectra. 
For \tsco{}, we used two CORALIE spectra obtained on 2007 March 18 and 
2007 March 25 with the 1.2\,m Euler Telescope at La~Silla. In the case of the $\beta$\,Cephei star \xcma{}, we 
made use of the average of a large number of time-resolved CORALIE exposures primarily obtained for a study of 
its pulsational properties (see Saesen et al.\ \cite{saesen}). The abundance results for this star are 
reported in Morel et al.\ (\cite{morel}), where complete details on the methodology used to derive the 
elemental abundances can also be found. 

A standard, iterative scheme is first used to derive the atmospheric parameters purely on spectroscopic 
grounds: $T_{\rm eff}$ is determined from the \ion{Si}{iii/iv} ionization balance, $\log g$ from fitting 
the collisionally-broadened wings of the Balmer lines and the microturbulent velocity $\xi$ from 
requiring the abundances yielded by the \ion{O}{ii} features to be independent of the line strength. As a 
final step, the projected rotational velocity is inferred by comparing the profiles of a set of isolated 
metal lines with a grid of rotationally-broadened synthetic spectra. Support for our temperature scale 
comes from the good agreement between the abundances yielded by different ions for elements other than Si. 
For \tsco{}, both \ion{C}{ii/iii} and \ion{N}{ii/iii} yield mean abundance values differing by at most 
0.11\,dex, which is typical of the line-to-line scatter ($\sigma_{\rm int}$). The same conclusion holds 
for \tcar{} in the case of \ion{N}{iii} $\lambda$4634, where the abundance is discrepant by only $\sim$2$\sigma_{\rm int}$
from the mean value given by the \ion{N}{ii} lines.
{
We infer a high microturbulent velocity in \tcar{} from the analysis of the \ion{O}{ii} lines
compared to the usual values found for early B-type dwarfs ($\xi$ = 12\,km\,s$^{-1}$).
A similar result, within the errors, is obtained using other species (i.e.\ N and Si).
}

\begin{table*}
\centering
\caption{Physical parameters of the programme stars (values for \xcma{} from Morel et al.\ \cite{morel}). 
} 
\label{tab:parameters}
\begin{tabular}{lr@{}c@{}lr@{}c@{}lr@{}c@{}l}
\hline
\hline
                         & \multicolumn{3}{c}{\xcma{}} & \multicolumn{3}{c}{\tsco{}} & \multicolumn{3}{c}{\tcar{}}            \\
\hline
Spectral type    & \multicolumn{3}{c}{B0.7 IV} & \multicolumn{3}{c}{B0.2 V} & \multicolumn{3}{c}{B0.2 V} \\
$T_{\rm eff}$ (K)        & 27\,500 & $\pm$ & 1000  & 31\,500 & $\pm$ & 1000 & 31\,000 & $\pm$ & 1000 \\
$\log g$ (cm s$^{-2}$)   & 3.75    & $\pm$ & 0.15  & 4.05    & $\pm$ & 0.15 & 4.20    & $\pm$ & 0.20 \\
$\xi$ (km\,s$^{-1}$)      & 6       & $\pm$ & 2     & 2       & $\pm$ & 2    & 12      & $\pm$ & 4    \\
$v \sin i$ (km\,s$^{-1}$) & 10      & $\pm$ & 2$^a$ & 8       & $\pm$ & 2    & 113     & $\pm$ & 8    \\ 
\hline
\end{tabular}
\begin{flushleft}
$^a$ This value is an upper limit, as the spectral lines are significantly broadened 
by pulsations in this $\beta$\,Cephei variable.
\end{flushleft}
\end{table*}

\begin{figure}
\centering
\includegraphics[width=0.45\textwidth]{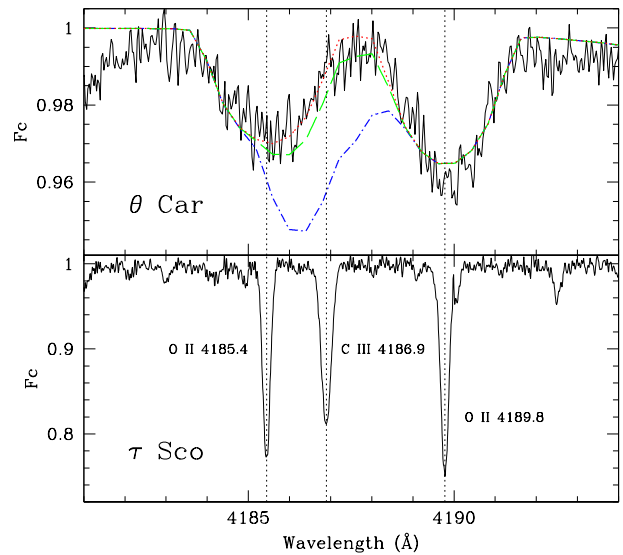}
\caption{{\em Upper panel:} Spectral synthesis in \tcar{} of the blend formed by 
\ion{O}{ii} $\lambda$4185.4 and \ion{C}{iii} $\lambda$4186.9. {\em Solid line:} observed spectrum, 
{\em dotted-short dashed line:} synthetic spectrum with a carbon abundance typical of B-type dwarfs 
($\log \epsilon$[C] = 8.20\,dex; see Table~\ref{tab:abundances}), {\em long-dashed line:} synthetic spectrum 
with the carbon abundance yielded by the \ion{C}{ii} $\lambda$4267 doublet ($\log \epsilon$[C] = 7.16\,dex), 
{\em dotted line:} synthetic spectrum with the best fit carbon abundance 
($\log \epsilon$[C] = 6.39\,dex). In the online version of this journal, the synthetic spectra are plotted 
as blue, green and red lines, respectively. A microturbulent velocity, $\xi$ = 12\,km\,s$^{-1}$, and an 
oxygen abundance, $\log \epsilon$(O) = 8.38\,dex, were assumed in all cases. All synthetic spectra have 
been convolved with a rotational broadening function with $v \sin i$ = 113\,km\,s$^{-1}$ 
(see Tables~\ref{tab:parameters} and \ref{tab:abundances}). {\em Lower panel:} the same spectral region 
in \tsco{} with the most prominent spectral lines identified. Note the different {\em y} scale.}
\label{fig:syn}
\end{figure}

The physical parameters for the three stars are provided in Table~\ref{tab:parameters}. An excellent agreement is found for 
\tsco{} with other modern NLTE studies (e.g., Kilian \cite{kilian}; Mokiem et al.\ \cite{mokiem}; 
Nieva \& Przybilla \cite{nieva}; Repolust et al.\ \cite{repolust}; Sim\'on-D\'{\i}az et 
al.\ \cite{simon_diaz}). The same parameters within the errors are obtained for all studied stars 
when analyzing our data with the unified code FASTWIND (Puls et al.\ \cite{puls}) and (semi-)automatic 
line-profile fitting techniques (Lefever \cite{lefever}). To our knowledge, the only previous abundance study dedicated 
to \tcar{} (Sch\"onberner et al.\ \cite{schonberner}) led to a $T_{\rm eff}$ value 1500\,K
higher than our estimate (as is also the case for \tsco{}), but this can be naturally 
understood as a consequence of their use of only slightly line-blanketed atmospheric models 
(Kudritzki \cite{kudritzki}).  

\begin{table*}
\centering
\caption{Mean NLTE abundances (on the scale in which $\log \epsilon$[H] = 12) and total 1$\sigma$-uncertainties
(values for \xcma{} from Morel et al.\ \cite{morel}). The number of spectral lines used 
is given in brackets. A blank indicates that the abundance of a given element could not be determined. 
For comparison purposes, we provide in the last column the typical values found for OB dwarfs in the 
solar neighbourhood (Daflon \& Cunha \cite{daflon_cunha}; Gummersbach et al.\ \cite{gummersbach}; 
Kilian-Montenbruck et al.\ \cite{kilian_montenbruck}). We define [N/C] and [N/O] as 
$\log$[$\epsilon$(N)/$\epsilon$(C)] and $\log$[$\epsilon$(N)/$\epsilon$(O)], respectively.}
\label{tab:abundances}
\begin{tabular}{lr@{}c@{}lr@{}c@{}lr@{}c@{}lc}
\hline
\hline
                    & \multicolumn{3}{c}{\xcma{}} & \multicolumn{3}{c}{\tsco{}} & \multicolumn{3}{c}{\tcar{}} & OB stars\\
\hline
He/H                & 0.098 & $\pm$ & 0.017 (10) & 0.085 & $\pm$ & 0.027 (9) & 0.083 & $\pm$ & 0.028 (5)    &  0.10$^a$ \\  
$\log \epsilon$(C)  & 8.18  & $\pm$ & 0.12 (9)   & 8.19  & $\pm$ & 0.14 (15) & 7.16  & $\pm$ & 0.46 (1)$^b$ &  $\sim$8.2 \\    
$\log \epsilon$(N)  & 8.00  & $\pm$ & 0.17 (34)  & 8.15  & $\pm$ & 0.20 (35) & 8.56  & $\pm$ & 0.27 (12)    &  $\sim$7.6 \\    
$\log \epsilon$(O)  & 8.59  & $\pm$ & 0.17 (34)  & 8.62  & $\pm$ & 0.20 (42) & 8.38  & $\pm$ & 0.22 (7)     &  $\sim$8.5 \\    
$\log \epsilon$(Mg) & 7.37  & $\pm$ & 0.20 (1)   & 7.45  & $\pm$ & 0.09 (2)  &       &       &              &  $\sim$7.4 \\    
$\log \epsilon$(Al) & 6.16  & $\pm$ & 0.22 (4)   & 6.31  & $\pm$ & 0.29 (3)  &       &       &              &  $\sim$6.1 \\    
$\log \epsilon$(Si) & 7.13  & $\pm$ & 0.21 (4)   & 7.24  & $\pm$ & 0.14 (9)  & 7.43  & $\pm$ & 0.23 (10)    &  $\sim$7.2 \\    
$\log \epsilon$(S)  & 6.99  & $\pm$ & 0.16 (2)   & 7.18  & $\pm$ & 0.28 (3)  & 7.32  & $\pm$ & 0.33 (1)     &  $\sim$7.2 \\    
$\log \epsilon$(Fe) & 7.30  & $\pm$ & 0.22 (32)  & 7.33  & $\pm$ & 0.31 (13) &       &       &              &  $\sim$7.3$^c$ \\
\hline		      			    	   		  			   
${\rm [N/C]}$       & --0.18  & $\pm$ & 0.21     & --0.04 & $\pm$ & 0.25     & +1.40 & $\pm$ & 0.53         &  $\sim$--0.6 \\  
${\rm [N/O]}$       & --0.59  & $\pm$ & 0.25     & --0.47 & $\pm$ & 0.29     & +0.18 & $\pm$ & 0.35         &  $\sim$--0.9 \\  
\hline
\end{tabular}
\begin{flushleft}
$^a$ From Lyubimkov et al.\ (\cite{lyubimkov}).\\
$^b$ Derived from a differential analysis of the \ion{C}{ii} $\lambda$4267 doublet with respect to \tsco{}  
(see text). \\
$^c$ Mean NLTE abundance from Morel et al.\ (\cite{morel}).
\end{flushleft}
\end{table*}

Curve-of-growth techniques were used to determine the abundances using the equivalent widths of a set 
of unblended lines in the relevant temperature range. The line list was constructed after careful 
inspection of a B0 spectral atlas based on the extensive line list of Kurucz \& Bell 
(\cite{kurucz_bell}).\footnote{Available online at: \\
{\tt http://www.lsw.uni-heidelberg.de/cgi-bin/websynspec.cgi}.}
In the case of the fast rotator \tcar{}, the sharp-lined spectrum of \tsco{} was used to further discard all diagnostic lines 
initially selected which could be significantly blended with nearby spectral features. No suitable 
carbon lines could be found owing to their extreme weakness (as already noted by Walborn \cite{walborn76})
and we followed in that case  these following complementary approaches:
\begin{itemize}
\item
Although the \ion{C}{ii} $\lambda$4267 doublet is unblended and reliably measurable, it is not 
appropriately modeled by our model atom and yields spuriously low abundances.\footnote{As is also 
the case for other NLTE line-formation codes (e.g.\ TLUSTY; Trundle et al.\ \cite{trundle}). See, 
however, Nieva \& Przybilla (\cite{nieva}) for a recent solution to this long-standing problem.} 
Nevertheless, the carbon abundance in \tcar{} can be roughly estimated through a {\em differential} 
analysis with respect to \tsco{} in view of their similar physical parameters (Table~\ref{tab:parameters}) 
and the lack of evidence for a difference in metallicity (the Si and S abundances are uncertain, 
but agree within the errors; Table~\ref{tab:abundances}).
We obtain an abundance in \tcar{} 0.98\,dex lower
than in \tsco{} from the analysis of \ion{C}{ii} $\lambda$4267. Scaling to the 
mean \ion{C}{ii} abundance derived for the latter translates into an absolute abundance in \tcar{}: 
$\log \epsilon$(C) = 7.16\,dex; 
\item
Alternatively, the C abundance can be inferred from a spectral synthesis of the blend formed by 
\ion{O}{ii} $\lambda$4185.4 and \ion{C}{iii} $\lambda$4186.9. We obtain an even lower best fit 
abundance: $\log \epsilon$(C) = 6.39\,dex (see Fig.~\ref{fig:syn}). 
\end{itemize}
Although we regard the carbon abundance yielded by \ion{C}{ii} $\lambda$4267 as more reliable, and will 
only consider this value in the following, the \ion{C}{iii} $\lambda$4186.9 analysis supports a very 
low carbon content and clearly rules out a solar value (as can be seen in Fig.~\ref{fig:syn}). 
The abundances are given in 
Table~\ref{tab:abundances} and are compared with the typical values found for nearby B-type dwarfs 
(e.g.\ Daflon \& Cunha \cite{daflon_cunha}; Gummersbach et al.\ \cite{gummersbach}; 
Kilian-Montenbruck et al.\ \cite{kilian_montenbruck}).
{
The quoted uncertainties are found by propagating the errors arising from the uncertainties
on the atmospheric parameters and the line-to-line scatter.
}

A one-order-of-magnitude nitrogen overabundance and carbon depletion is found in \tcar{}, but the oxygen 
abundance is roughly solar.
{
Sch\"onberner et al.\ (\cite{schonberner}) found $\rm [N/C]=+1.7$ and $\rm [N/O]=+0.3$,
which agree, within the errors, with our values.
}
Such abundance anomalies identify it as a member of the rare OBN class (Walborn \cite{walborn76})
and are consistent with its classification as a blue straggler (Schild \& Berthet \cite{schild}). 
Two main formation channels may account for the existence of this population in young open clusters. 
On the one hand, they have been claimed to result from the quasi-chemically 
homogeneous evolution of single stars initially rotating at a significant fraction of breakup 
velocity (Maeder \cite{maeder87}). However, a strong O depletion and 
He enrichment may be expected, which is not the case (we do not confirm the moderate helium enhancement, 
He/H = 0.17, reported by Sch\"onberner et al.\ \cite{schonberner}). Very high rotation rates are also 
required in this scenario. In the recent models of Meynet \& Maeder (\cite{meynet}), even a 15\,M$_{\sun}$ star with a 
rotational velocity as high as 300\,km\,s$^{-1}$ on the Zero Age Main Sequence (ZAMS) follows a normal redward evolution. 
Furthermore, complete mixing on the main sequence is much more likely to occur in low-metallicity 
environments because of the small loss of angular momentum resulting from the much lower mass-loss rates (e.g.\
Woosley \& Heger \cite{woosley}). On the other hand, there is strong observational evidence in dwarfs 
for a link between the OBN phenomenon and binarity (e.g.\ Levato et al.\ \cite{levato}). The high rotation 
rate and altered CN abundances would result in that case from the accretion during a past episode of 
mass transfer of incomplete CNO-cycle processed material from the initially more massive component. 
In the framework of this binary scenario, the solar helium abundance would indicate that complete 
mixing of the mass gainer has not taken place during the accretion process (Vanbeveren \& de Loore 
\cite{vanbeveren}).

A different abundance pattern is found in the single stars \xcma{} and \tsco{}. First, the carbon abundance 
is solar and, while an N excess is also detected, it is of much smaller amplitude (0.4--0.6\,dex).
This N overabundance is typical of the values already found for other slowly-rotating (magnetic) B-type dwarfs 
and may require another physical explanation (Morel et al.\ \cite{morel08}).

\begin{figure*}
\centering
\includegraphics[width=0.45\textwidth]{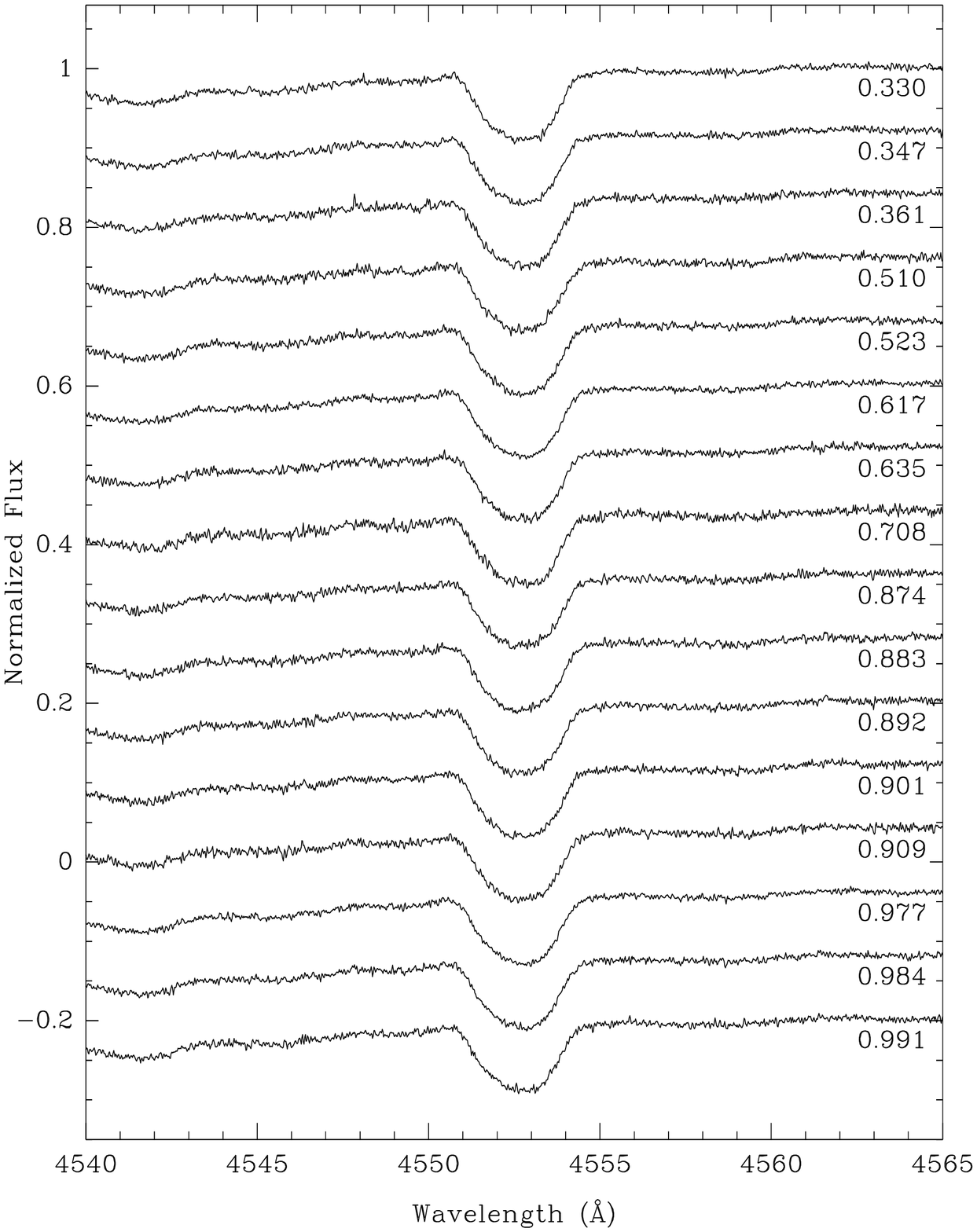}
\includegraphics[width=0.45\textwidth]{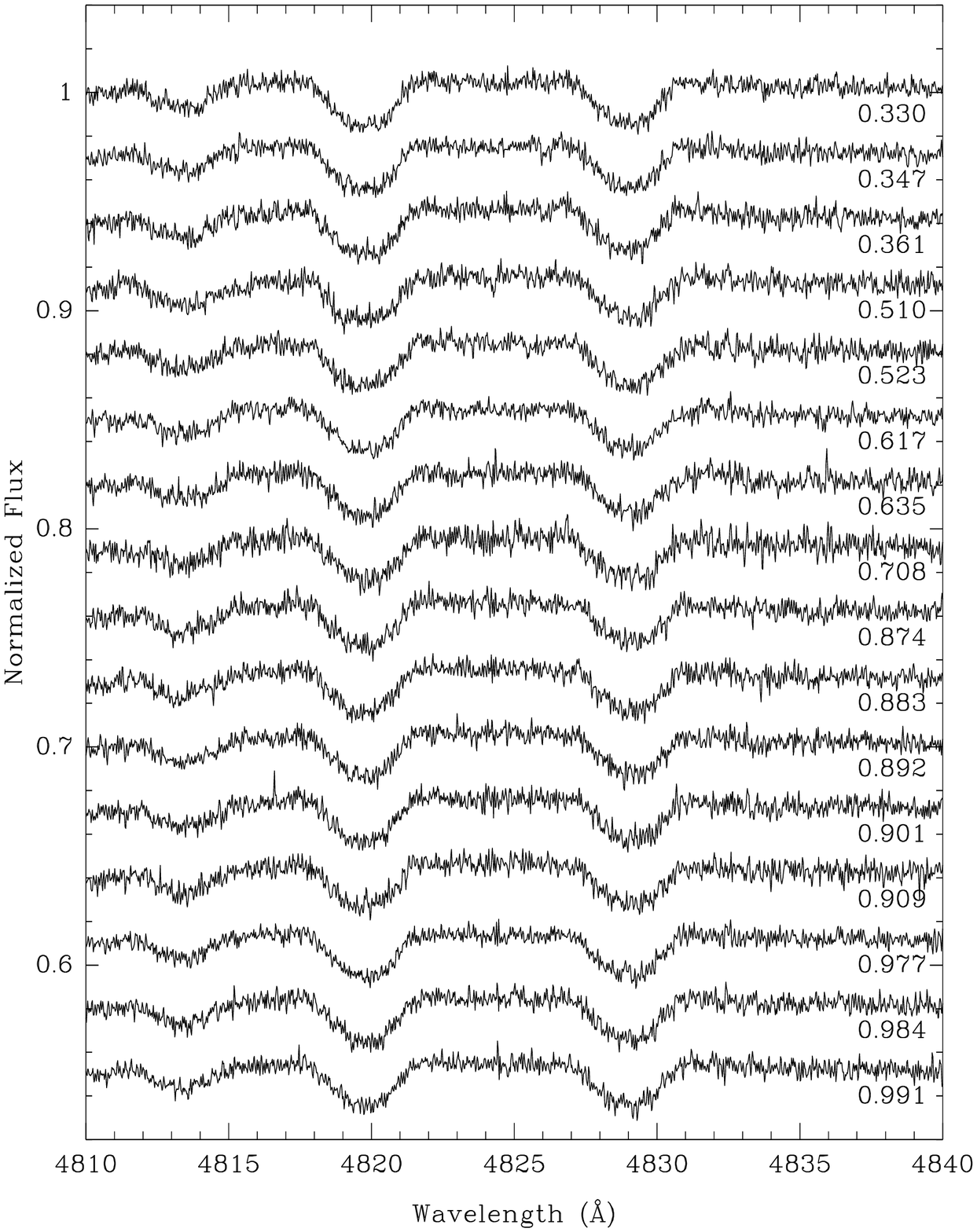}
\caption{
Weak spectral line variability of \ion{Si}{iii} lines in \tcar{}.
{\em Left:}
\ion{Si}{iii} at $\lambda$4552.6.
{\em Right:}
\ion{Si}{iii} at $\lambda$4819.6/$\lambda$4819.7 and $\lambda$4829.0.
The normalized spectra are from 16 different orbital phases.
Each spectrum is an average of three to five CORALIE spectra obtained close in time.
The spectra are shifted vertically for display purposes.
}
\label{fig:4553+4820}
\end{figure*}

In our CORALIE spectra, the \ion{Si}{iii} absorption lines show weak variability over
the orbital period (Fig.~\ref{fig:4553+4820}).
The \ion{Si}{iii} line at $\lambda$4552.6 displays an
asymmetric profile at phases 0.71 and 0.98--0.99, while it is rather symmetric at phases 0.87--0.91.
The blend of the two \ion{Si}{iii} lines at $\lambda$4819.6/$\lambda$4819.7 seems to have a sharper core
(like a triangle) in phases  0.6--0.9
and is more round or square shaped in the phase range 0.3--0.5.
The detected spectral variability could be caused by the presence of surface Si spots,
by pulsations, or by the presence of the companion.  
Our spectral material is, however,
insufficient to test the origin of this variability.
We believe that also the \ion{He}{i} lines 
are weakly variable, but a detailed study of the presence 
of He and Si spots on the surface of \tcar{} requires much higher spectral 
resolution and higher S/N spectra. Interestingly, a similar low-level variability
in He lines has been 
detected in the spectra of the SB1 HD\,191612 of spectral type Of?p
(Naz{\'e} et al.\ \cite{naze}), for which a magnetic field has been detected by 
Donati et al.\ (\cite{donati06a}).

\section{Spectropolarimetric observations}
\label{sect:magfield}

The only search for a mean longitudinal magnetic field in \tcar{} before our FORS\,1 observations
has been conducted by Borra \& Landstreet (\cite{borra_land})
using a photoelectric Pockels cell polarimeter for the measurement of the circular polarization
in the wings of the H$\beta$ line. 
Measuring the mean longitudinal magnetic field is currently the standard method for searching magnetic 
fields in different types of stars.
The mean longitudinal magnetic field is the average over the visible stellar hemisphere of the component 
of the magnetic field vector along the line of sight.
Due to the sensitivity to aspect, it provides essential constraints for all models of geometry and gives 
a detailed structure of the magnetic fields of these stars.

\tcar{} was observed by Borra \& Landstreet (\cite{borra_land}) on four nights with exposure times
between 40 and 64\,min, but no field was 
detected in spite of rather small standard deviations of the order of 50--65\,G. A presence of 
variable emission in 
the H$\beta$ line, which could possibly dilute the Zeeman polarization signal, was not reported by previous 
spectroscopic studies, hence the non-presence of a magnetic field seemed to be justified.

\begin{figure}
\centering
\includegraphics[width=0.35\textwidth]{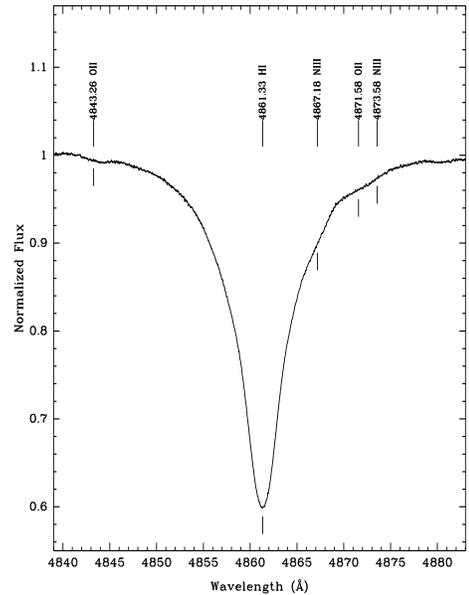}
\caption{
The H$\beta$ line for \tcar{} with \ion{N}{iii} blends in the red wing.
}
\label{fig:hbeta}
\end{figure}

The careful inspection of our FEROS spectrum and CORALIE spectra of \tcar{} does not reveal any 
emission-like feature 
in the core of the H$\beta$ line. However, we detect a clear blend contribution in the red wing 
of this line, mainly due to \ion{N}{iii} lines (Fig.~\ref{fig:hbeta}). As this element is found 
overabundant in \tcar{} by a one-order magnitude, the blend can in principle affect the previous
magnetic field measurements which made use of circular polarization present in the wings of the
H$\beta$ line.

The first observations of \tcar{} with FORS\,1 at the VLT in polarimetric mode using GRISM 600B 
were obtained in 2004 January in the framework of our survey of magnetic fields 
in Ap and Bp stars with accurate Hipparcos parallaxes (Hubrig et al.\ \cite{hubrig07}).
The observations consisted of four series 
of two exposures with the retarder waveplate oriented at two different angles ($\alpha$ = +45$^\circ$ and $-$45$^\circ$).
During this run we used the GRISM 600B to cover the whole spectral region from H$\beta$ to 
the Balmer jump.
With the narrowest available slit width of 0$\farcs$4 
we could achieve a spectral resolving power of $R \approx 2000$. The highest signal-to-noise ratio at
$\sim$4300\,\AA{} was about 650. 
Quite puzzling, the magnetic field measurements on the Stokes V spectra obtained with an exposure time 
of 0.2\,s showed a change of the field  polarity after the first series
(which took 2.5\,min) from a positive magnetic field to a negative magnetic field. 
The polarity change could, however, not obviously be explained by any instrumental effect since no change 
of polarity on short time scales was observed in the 
magnetic Ap star HD\,99563, observed in the same night just after \tcar{}.
To solve the puzzle, we re-observed this star in 2007 March at higher spectral resolution and at 
higher S/N (S/N$\sim$1200) with the 
GRISM 1200B in the spectral region from 3805\,\AA{} to 4960\,\AA{} at a spectral resolving power of $R \approx 4000$.
To monitor the behaviour of the magnetic field over at least a part of 
the stellar surface we carried out a time series of 
exposures with short integration time over the time span of $\sim$1.2\,h. In all, we
obtained 26 series of magnetic field measurements with exposure times of 0.8\,s and 1.1\,s. 
An additional star with a 
well-defined strong longitudinal field, HD\,96440,  was selected as a standard star to check that the
instrument was functioning properly. This star has a mean longitudinal
magnetic field that varies about a mean value of $\sim-1900$\,G with a low amplitude ($\sim$160\,G
peak-to-peak) over a period of 2800\,d (Mathys et al., in preparation).

\begin{table*}
\centering
\caption{Individual measurements of the mean longitudinal magnetic field for \tcar{} for the observations
in 2004 and 2007.}
\label{tab:results}
\begin{tabular}{cr|r@{}c@{}r|r@{}c@{}r|r@{}c@{}r|c}
\hline
\hline
\multicolumn{1}{c}{MJD} &
\multicolumn{1}{c}{time} &
\multicolumn{1}{|c}{$\left< B_z \right>$} &
\multicolumn{1}{c}{$\pm$} &
\multicolumn{1}{c}{$\sigma_B$} &
\multicolumn{1}{|c}{$\left< B_z \right>$} &
\multicolumn{1}{c}{$\pm$} &
\multicolumn{1}{c}{$\sigma_B$} &
\multicolumn{1}{|c}{$\left< B_z \right>$} &
\multicolumn{1}{c}{$\pm$} &
\multicolumn{1}{c}{$\sigma_B$} &
\multicolumn{1}{|c}{phase} \\
\multicolumn{1}{c}{} &
\multicolumn{1}{c}{[s]} &
\multicolumn{1}{|c}{} &
\multicolumn{1}{c}{[G]} &
\multicolumn{1}{c}{} &
\multicolumn{1}{|c}{} &
\multicolumn{1}{c}{[G]} &
\multicolumn{1}{c}{} &
\multicolumn{1}{|c}{} &
\multicolumn{1}{c}{[G]} &
\multicolumn{1}{c}{} &
\multicolumn{1}{|c}{} \\
\multicolumn{1}{c}{} &
\multicolumn{1}{c}{} &
\multicolumn{3}{|c}{set 1} &
\multicolumn{3}{|c}{set 2} &
\multicolumn{3}{|c}{set 3} &
\multicolumn{1}{|c}{} \\
\hline
53012.22798 & 0.0 &    247 & $\pm$ & 130 &    692 & $\pm$ & 184 &    322 & $\pm$ & 211 & \\
53012.22973 & 151.2 & $-$136 & $\pm$ & 131 & $-$267 & $\pm$ & 186 & $-$584 & $\pm$ & 216 & \\
53012.23149 & 303.3  & $-$219 & $\pm$ & 125 & $-$238 & $\pm$ & 178 & $-$342 & $\pm$ & 206 & \\
53012.23324 & 454.5 & $-$122 & $\pm$ & 142 & $-$176 & $\pm$ & 205 & $-$242 & $\pm$ & 223 & \\
\hline
54181.00814 & 0.0 & $-$46 & $\pm$ & 52 & $-$223 & $\pm$ & 92 & $-$47 & $\pm$ & 112 & 0.000 \\
54181.01007 & 166.1 & 49 & $\pm$ & 62 & 148 & $\pm$ & 78 & 113 & $\pm$ & 98 & 0.316 \\
54181.01214 & 345.3 & $-$106 & $\pm$ & 55 & $-$25 & $\pm$ & 77 & 18 & $\pm$ & 96 & 0.656 \\
54181.01421 & 524.1 & $-$57 & $\pm$ & 45 & 69 & $\pm$ & 83 & 56 & $\pm$ & 101 & 0.996 \\
54181.01613 & 690.2 & 57 & $\pm$ & 55 & 26 & $\pm$ & 90 & 1 & $\pm$ & 110 & 0.311 \\
54181.01805 & 856.3 & 34 & $\pm$ & 47 & $-$38 & $\pm$ & 78 & $-$226 & $\pm$ & 97 & 0.627 \\
54181.01997 & 1022.2 & 53 & $\pm$ & 47 & 84 & $\pm$ & 88 & 100 & $\pm$ & 109 & 0.942 \\
54181.02190 & 1188.2 & 48 & $\pm$ & 50 & 136 & $\pm$ & 82 & 207 & $\pm$ & 102 & 0.258 \\
54181.02382 & 1354.2 & $-$49 & $\pm$ & 58 & $-$51 & $\pm$ & 95 & $-$185 & $\pm$ & 120 & 0.573 \\
54181.02574 & 1520.3 & 266 & $\pm$ & 52 & 220 & $\pm$ & 85 & 244 & $\pm$ & 108 & 0.889 \\
54181.02766 & 1686.5 & 53 & $\pm$ & 57 & 85 & $\pm$ & 93 & $-$37 & $\pm$ & 118 & 0.204 \\
54181.02959 & 1853.0 & $-$81 & $\pm$ & 60 & $-$239 & $\pm$ & 100 & $-$305 & $\pm$ & 124 & 0.521 \\
54181.03153 & 2020.3 & 64 & $\pm$ & 50 & 32 & $\pm$ & 82 & 219 & $\pm$ & 104 & 0.839 \\
54181.03345 & 2186.6 & 26 & $\pm$ & 51 & $-$3 & $\pm$ & 84 & $-$5 & $\pm$ & 105 & 0.155 \\
54181.03537 & 2352.5 & $-$131 & $\pm$ & 72 & $-$245 & $\pm$ & 92 & $-$459 & $\pm$ & 118 & 0.470 \\
54181.03730 & 2519.2 & 182 & $\pm$ & 65 & 176 & $\pm$ & 105 & 423 & $\pm$ & 132 & 0.786 \\
54181.03960 & 2717.6 & 81 & $\pm$ & 45 & $-$77 & $\pm$ & 75 & 58 & $\pm$ & 93 & 0.164 \\
54181.04154 & 2885.4 & $-$210 & $\pm$ & 52 & $-$202 & $\pm$ & 87 & $-$287 & $\pm$ & 110 & 0.482 \\
54181.04348 & 3053.4 & 37 & $\pm$ & 50 & $-$92 & $\pm$ & 80 & $-$86 & $\pm$ & 103 & 0.802 \\
54181.04542 & 3220.4 & 118 & $\pm$ & 50 & 45 & $\pm$ & 84 & 205 & $\pm$ & 105 & 0.119 \\
54181.04736 & 3388.5 & $-$161 & $\pm$ & 62 & 98 & $\pm$ & 102 & $-$180 & $\pm$ & 129 & 0.438 \\
54181.04931 & 3556.5 & $-$112 & $\pm$ & 57 & 123 & $\pm$ & 93 & $-$82 & $\pm$ & 117 & 0.757 \\
54181.05125 & 3724.7 & 79 & $\pm$ & 50 & $-$53 & $\pm$ & 82 & $-$46 & $\pm$ & 105 & 0.077 \\
54181.05320 & 3892.7 & $-$159 & $\pm$ & 65 & $-$462 & $\pm$ & 98 & $-$394 & $\pm$ & 125 & 0.396 \\
54181.05514 & 4060.9 & $-$114 & $\pm$ & 57 & $-$214 & $\pm$ & 95 & $-$269 & $\pm$ & 122 & 0.716 \\
54181.05709 & 4228.7 & $-$36 & $\pm$ & 57 & 38 & $\pm$ & 100 & 151 & $\pm$ & 128 & 0.034 \\
\hline
\end{tabular}
\end{table*}

\begin{table}
\centering
\caption{Individual measurements of the mean longitudinal magnetic field for the strongly magnetic star HD\,94660.} 
\label{tab:results2}
\begin{tabular}{c|r@{}c@{}r|r@{}c@{}r}
\hline
\hline
\multicolumn{1}{c}{MJD} &
\multicolumn{1}{|c}{$\left< B_z \right>$} &
\multicolumn{1}{c}{$\pm$} &
\multicolumn{1}{c}{$\sigma_B$} &
\multicolumn{1}{|c}{$\left< B_z \right>$} &
\multicolumn{1}{c}{$\pm$} &
\multicolumn{1}{c}{$\sigma_B$} \\
\multicolumn{1}{c}{} &
\multicolumn{1}{|c}{} &
\multicolumn{1}{c}{[G]} &
\multicolumn{1}{c}{} &
\multicolumn{1}{|c}{} &
\multicolumn{1}{c}{[G]} &
\multicolumn{1}{c}{} \\
\multicolumn{1}{c}{} &
\multicolumn{3}{|c}{set 1} &
\multicolumn{3}{|c}{set 2} \\
\hline
54181.03960 & $-$1534 & $\pm$ & 30 & $-$1919 & $\pm$ & 56 \\
54181.04154 & $-$1624 & $\pm$ & 32 & $-$1861 & $\pm$ & 60 \\
54181.04348 & $-$1688 & $\pm$ & 34 & $-$1889 & $\pm$ & 62 \\
54181.04542 & $-$1596 & $\pm$ & 33 & $-$1908 & $\pm$ & 62 \\
\hline
\end{tabular}
\end{table}

The individual measurements of the longitudinal magnetic field 
obtained in 2004 and 2007 are presented in Table~\ref{tab:results}.
The columns are, in order, the modified Julian date 
of the middle of the exposures, the time elapsed from the start of the first exposure, the mean longitudinal 
magnetic field $\left<B_z\right>$ 
measured on Balmer lines and metal lines (set\,1), the mean longitudinal magnetic field measurements 
restricted to the wavelength regions containing hydrogen Balmer lines (set\,2), and 
the field measured on Balmer lines with the exception of the H$\beta$ line (set\,3). The last column 
gives the corresponding phase assuming a variability period of about 8.8~minutes.
The measurements for the strongly magnetic standard star HD\,94660 are presented in Table~\ref{tab:results2}.

\begin{figure}
\centering
\includegraphics[height=0.45\textwidth,angle=270]{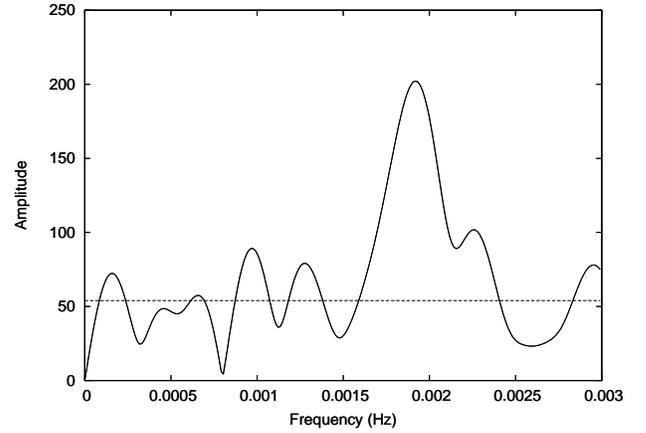}
\caption{
Fourier spectrum for the magnetic field data derived from hydrogen lines without H$\beta$.
The horizontal line gives the noise level.} 
\label{fig:amplit}
\end{figure}

\begin{figure}
\centering
\includegraphics[height=0.45\textwidth,angle=270]{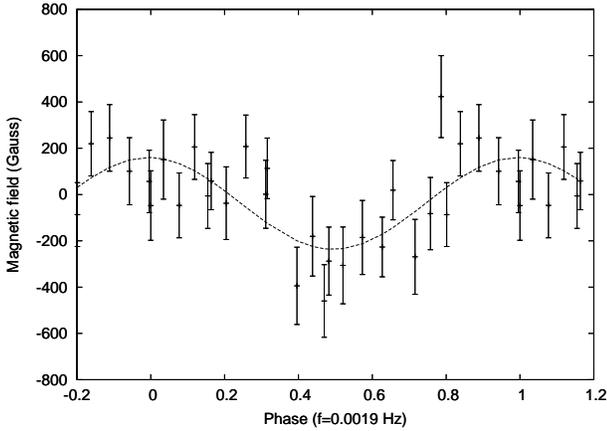}
\caption{
Phase diagram for the same dataset as in Fig.~\ref{fig:amplit} for the best 
candidate frequency of 0.0019\,Hz, i.e.\ a period of about 8.8~minutes.
}
\label{fig:phase}
\end{figure}

Our observations obtained in 2007 March confirm the previous finding of the magnetic field variations on 
a short time scale. The observed magnetic field changes several times from positive to negative values 
over the observing time span. However, only a few measurements are of the order of 
a few hundred Gauss at a significance level of more than 3$\sigma$. 
On the other hand, the amplitude of the variability of the magnetic field  of our 
strongly magnetic standard star HD\,94660 accounts only for $\sim$77\,G.
To find a potential period of variations of the magnetic field in \tcar{} we performed a Fourier analysis of 
all three sets of magnetic field determinations.
For the Fourier analysis we used Period04
(Lenz \& Breger \cite{lenz_breger}) and searched for the best candidate frequency in the 
range [0,$f_{\rm Nyq}$]\,Hz, where $f_{\rm Nyq}$ is the Nyquist frequency of 0.003\,Hz.
As a result, we find that the frequency 0.0019\,Hz is present in all three sets but with the highest S/N 
level in set\,3.
For set\,1 (with measurements of Balmer and metal lines) and for set\,3 (with the H$\beta$ line 
excluded), the highest peaks are at about 0.0019\,Hz, 
at a 3.0$\sigma$ level and 3.8$\sigma$ level, respectively.
For set\,2, which includes the measurements on all available Balmer lines, only the second highest peak is 
at the same frequency, 0.0019\,Hz, at a 2.5$\sigma$ level.
The Fourier spectrum  for set\,3 is shown in  Fig.~\ref{fig:amplit} with an amplitude of 
the magnetic field variation of 202$\pm$43\,G.
The corresponding phase diagram 
is shown in Fig.~\ref{fig:phase}.

\section{Discussion}
\label{sect:disc}

\subsection{Binarity and evolutionary history}

\begin{figure}
\centering
\includegraphics[height=0.48\textwidth,angle=270]{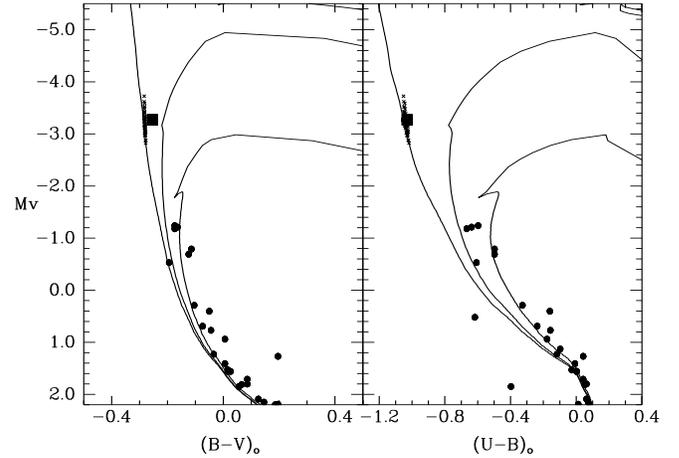}
\caption{Color-Magnitude diagram of IC\,2602. 
Small crosses mark the position of main sequence stars with $T_{\rm eff}$ = 30\,000--32\,000\,K
and $\log g$ = 4.0--4.4, according to theoretical models.
Solid lines represent isochrones for $\log\tau$ = 3.0, 7.5, and 8.0.
The photometric position of 
\tcar{} is marked with a filled square and other IC\,2602 members  with filled circles.
Photometric data and cluster parameters [$(m-M)$ = 6.1, $E(B-V)$ = 0.024]
were taken from the WEBDA database (Mermilliod \cite{mermilliod}).
}
\label{fig:cmd}
\end{figure}

We have presented in Sect.~\ref{sect:binarity} the first high 
quality radial velocity curve for \tcar{}.
It is well fitted by a Keplerian orbit with eccentricity $e$ = 0.127,
leaving no doubt about the binary origin of these variations. 
The improved determination of the orbital parameters allows us to make 
a more sophisticated interpretation of the nature of this system
and to review the possibility of a mass transfer between primary and secondary
as the cause of the observed chemical anomalies.
The physical parameters of \tcar{} determined in  Sect.~\ref{sect:abundances} suggest 
that it is located in the H-R diagram close to the ZAMS.
According to the Geneva stellar models (Schaller et al.\ \cite{schaller}; Lejeune \& Schaerer \cite{lejeune_schaerer})
for solar abundances, the observed values of $T_{\rm eff}$ = 31\,000\,K
and $\log g$ = 4.2 correspond to a star with mass $M_1$ = 15.25\,M$_{\sun}$, radius
$R_1$ = 5.1\,R$_{\sun}$, and  age $\log\tau$ = 6.0.
Considering a 1$\sigma$ uncertainty in $\log g$ and $T_{\rm eff}$, 
we obtain $\log\tau \le 6.7$. This age value is
considerably lower than the accepted age for the cluster IC\,2602
($\log\tau = 7.83$, Kharchenko et al.\ \cite{kharchenko};
$\log\tau = 7.63$, Eggen \cite{eggen};
$\log\tau = 7.32$, Whiteoak \cite{whiteoak};
$\log\tau = 7.16$, Hill \& Perry \cite{hill_perry}),
consistent with the classification of
this star as a blue straggler.
We note that the absolute magnitude and intrinsic colors 
interpolated in the same
theoretical grid are in agreement with the position of \tcar{} in
the Color--Magnitude diagram of the cluster IC\,2602 (see Fig.~\ref{fig:cmd}).

\begin{figure}
\centering
\includegraphics[width=0.35\textwidth]{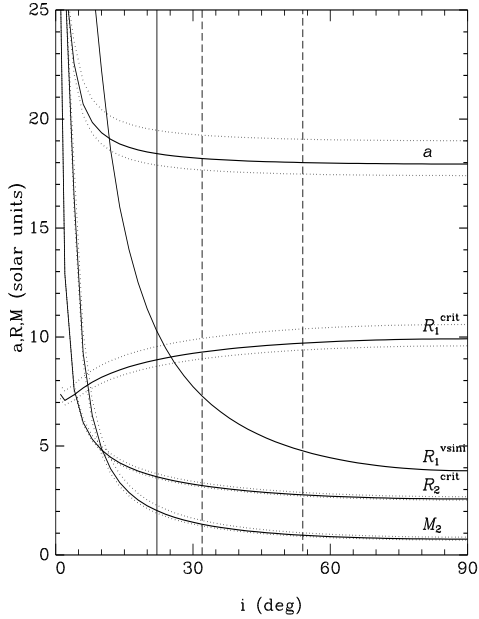}
\caption{Physical  parameters as a function of the orbital inclination:
orbital semiaxis $a$, critical radius for the primary $R_1^{\rm crit}$
and the secondary $R_2^{\rm crit}$, primary radius derived from
the rotational velocity $R_1^{\rm vsini}$, and mass of the secondary star $M_2$.
The full lines correspond to $M_1$ = 15.25\,M$_{\sun}$ and the dotted lines to $M_1$ = 13.9\,M$_{\sun}$ and $M_1$ = 18.2\,M$_{\sun}$.}
\label{fig:param}
\end{figure}

The uncertainty in the stellar mass and radius interpolated in the stellar model grid
was estimated from the adopted errors for $T_{\rm eff}$ and $\log g$.
The primary mass ranges from 13.9\,M$_{\sun}$ (corresponding to a ZAMS star with
$T_{\rm eff} = 30\,000$\,K) to 18.2\,M$_{\sun}$ (corresponding to $T_{\rm eff} = 32\,000$\,K and 
$\log g = 4.0$).
From the estimated  primary mass, relevant information for
the system can be derived from the radial velocity curve, even when
the orbital inclination is in principle unknown.
We show in Fig.~\ref{fig:param}  the absolute value of the orbital
semiaxis and the mass of the secondary star 
as a function of the orbital inclination. 
For each parameter plotted in this figure the solid line corresponds to 
the solution with $M_1 = 15.25$\,M$_{\sun}$,
while solutions with 13.9\,M$_{\sun}$ and 18.2\,M$_{\sun}$ are plotted with dotted lines.

To evaluate the occurrence of a mass exchange in the
system, we calculated the volume radius of the Roche lobe at periastron
($R_1^{\rm crit}$ and $R_2^{\rm crit}$), which are plotted in Fig.~\ref{fig:param}.
From this figure it is clear that these parameters are not strongly dependent on the
orbital inclination unless the latter is low.
However, a low inclination can be ruled out for the following reasons.
On the one hand, as argued by Walborn (\cite{walborn79}), the absence of a pronounced
emission at H$\alpha$ indicates that the equatorial velocity is
lower than 300\,km\,s$^{-1}$ and consequently, from the  observed $v \sin i$ value
we obtain $i \geq 22^\circ$. 
This value is presented by the solid vertical line in Fig.~\ref{fig:param}
and corresponds to an upper limit of about 2.0\,M$_{\sun}$ for the
secondary component.
On the other hand, if we assume that the rotation is synchronized with
the orbital motion at periastron, the relatively small radius deduced 
from the spectroscopic value of $\log g$ is compatible with the observed 
$v \sin i$ only for high inclinations.
For the case of synchronous rotation we derive a rotational period of 1.692$\pm$0.007\,d
and hence $R_1\sin i$ = 3.78$\pm$0.27\,R$_{\sun}$.
The adopted range for $M_1$ (13.9--18.2\,M$_{\sun}$) corresponds to
$R_1$ = 4.7--7.1\,R$_{\sun}$, and therefore we obtain $i = 32^\circ - 54^\circ$.
These limits are marked with two  dashed vertical lines
in Fig.~\ref{fig:param}.
Interestingly, this rotational period of 1.692$\pm$0.007\,d  is in good agreement with
the 1.779\,d period discovered by Walborn (\cite{walborn79}) who studied intensity variations of
\ion{Si}{IV} $\lambda$4089.

The radius derived from $v \sin i$ assuming synchronization would coincide
with the critical radius for $i \approx 25^\circ$.
In this case $R_1 \approx 9.1\,$R$_{\sun}$
and $\log g \approx 3.7$, which is lower than the spectroscopic value by  2.5$\sigma$.
Therefore, we expect that, at present, the primary star is probably not filling
its Roche lobe, although it is close to that point. 

A star of 15.25\,M$_{\sun}$ (13.9--18.5\,M$_{\sun}$) with a present radius of 5.1\,R$_{\sun}$ 
(4.7--7.1\,R$_{\sun}$) has an evolutionary age of 1.0\,Myr (0--4.3\,Myr) and, evolving as
a single star, would have at the Terminal Age Main Sequence (TAMS) a radius of about 13.6\,R$_{\sun}$ (12.6--16.3\,R$_{\sun}$), 
which is significantly larger than the critical radius. 
Thus, we can expect mass-transfer to occur from the primary to the secondary
during the main-sequence stage, at an evolutionary age of about 11\,Myr.

Since  \tcar{} is a bona-fide blue straggler of IC\,2602, it is very likely that the system 
has suffered in the past a mass-transfer in the opposite direction. 
Eggen \& Iben (\cite{eggen_iben})
have discussed extensively this possibility as an explanation for its
abnormal position in the Color-Magnitude diagram.
The apparent evolutionary youth of \tcar{} and its peculiar chemical abundances are best
understood if the present primary star was originally the less-massive component
and has accreted mass transfered from the companion.  
Its original mass could be less than half of the total mass of the system, similar
to other stars near the turnoff point in the cluster.
After considerable mass-transfer, it would appear as a non-evolved massive star
above the cluster turnoff and near the ZAMS.
As was noted by Eggen \& Iben (\cite{eggen_iben}), during the evolution of this binary system
an important loss of angular momentum has
taken place, making it possible to achieve a system with such a short period and low
mass-ratio.
The orbital angular momentum of a binary system can be written as 
$$
J_{\rm orb} = \frac{2 \pi}{P} M a^2 \frac{q}{(1+q)^2} \sqrt{1-e^2}
$$
while for the rotational momentum of each star we have
$$
J_{rot} = \frac{2 \pi}{P_{rot}} \beta^2 M R^2
$$
where $\beta$ is the radius of gyration, which is typically 0.25 for a
main-sequence star (Claret \& Gim\'enez \cite{claret_gimenez}).
In any case the rotational contribution to the total angular momentum is
small.
We can use these expressions to calculate the  angular momentum of the system  and compare 
the present value with that at the moment when the
mass-ratio was close to unity. As illustration, let us assume that the original
primary star (presently the secondary) with 9\,$M_{\sun}$ filled its Roche lobe 
when it was still in the main-sequence with a radius of the order of 7--9\,$R_{\sun}$.
Therefore, at that time the separation between the stars was about 20--25\,$R_{\sun}$.
Assuming a conservative mass transfer and 
adopting for the present configuration  $M_1$ = 15\,$M_{\sun}$, $M_2$ = 1.0\,$M_{\sun}$,
and a = 19\,$R_{\sun}$,  we can estimate  that the
present angular momentum is about 0.22--0.25 of the original value. 
In conclusion, most of the angular momentum has been lost.

\subsection{Abundances and magnetic field}


The binary and blue straggler nature of \tcar{} leads us to relate the observed chemical anomalies
to the past history of the binary system.
First, mass transfer of chemically processed material from the initially more massive component
has dramatically altered the CNO surface abundances.
Second, the spun up phase that ensued may have further enhanced nitrogen
and depleted carbon in \tcar{} as a result of rotational mixing
(see e.g.\ Langer et al.\ \cite{langer}).
The tendency of magnetic B stars to show a higher incidence
of an N excess compared to stars with no field detection (Morel et al.\ \cite{morel08})
also indicates that magnetic phenomena could play a role.
The results of the previous studies, along with the fact that little is known about the magnetic properties
of hot, massive stars, has motivated our search for a magnetic field in \tcar{}.

There are only six massive stars with detected magnetic fields that are hotter than \tcar{}:
$\theta^1$\,Ori~C, HD\,191612, HD\,155806, HD\,148937, 9\,Sgr, and \tsco{}.
The star HD\,191612 with a rotation period of 538\,d was observed 
with ESPaDOnS only during four consecutive nights (Donati et al.\ \cite{donati06a}). 
Among the published magnetic field measurements 
of $\theta^1$\,Ori~C, only four measurements show significance at the 3$\sigma$ level (Wade et al.\ \cite{wade06}).
HD\,155806 and HD\,148937 were observed only once and 9\,Sgr three times
(Hubrig et al.\ \cite{hubrig07}; Hubrig et al., in preparation). 
The sixth star, \tsco{},
with physical parameters very similar to those of \tcar{}, 
was studied over the rotational period of 41\,d.
It possesses the weakest mean longitudinal magnetic field with a maximum value of 88\,G and a very 
complicated geometry.
The structure of its magnetic field topology features in particular a significantly warped torus 
of closed magnetic loops encircling the star and additional smaller networks of closed field lines
(Donati et al.\ \cite{donati06b}).

In our magnetic field study of \tcar{}, only very few measurements were 
achieved at a significance level of 3$\sigma$ and their sporadic appearance is difficult to 
explain in the framework of the presence of a global large-scale organized magnetic field.
On the other hand, the presence of a complex structure of the magnetic field in \tcar{} may 
hypothetically be
possible in an analogous manner to the magnetic field topology of \tsco{}.
The detected periodicity of the order of 8.8\,min in the dataset of measurements carried out 
on hydrogen Balmer lines with the exclusion of H$\beta$ is surprising, and if it is not spurious, 
its discovery would give rise to the important question whether the presence of pulsations could cause
such a periodicity. No studies of short-time pulsations exist for \tcar{} so far.
B0 main-sequence stars are expected to pulsate in low radial order p-  
and g-modes with periods of the order of hours (typically 3 to 8  
hours). For such massive stars, periods of the order of a few minutes  
would correspond to high radial order p-modes, but non-adiabatic codes  
do not predict their excitation. However, since current models  
do not take into account the presence of a magnetic field, one cannot  
exclude the possibility that magnetic fields would favour the  
excitation of these types of modes in analogy to roAp stars which do  
possess a magnetic field and pulsate in high radial order modes of the  
order of a few minutes.
Our data with the CORALIE spectrograph were taken typically every 7~minutes.
This leads to a sampling rate which is too low to allow us to detect 
variations of spectral line profiles on short time scales.

Since the mechanism of the generation and the maintenance of magnetic fields 
in massive stars is not well understood yet, we are cautious in drawing any 
conclusion on the presence and the behaviour of a magnetic field in \tcar{}. 
Only additional time-resolved magnetic field observations will tell us about the 
presence and the structure of the magnetic field geometry of \tcar{} and will help to discriminate 
among the hypotheses described above.

\begin{acknowledgements}
T.~M. acknowledges financial support from the Research Council of Leuven University through grant 
GOA/2003/04. We are very grateful to C.~Aerts and M.~Desmet for acquiring the CORALIE spectra of \tsco{}, 
to V.~Elkin and P.~North for the FEROS spectra of \tcar{}, and to M.~E.~Veramendi for acquiring the CASLEO spectra.
We also wish to thank L.~Decin and M.~F.~Nieva for 
their help on the construction of the model grid at high temperatures,
K.~Lefever for communicating her results prior to publication,
and Y.~Naz\'e  and the referee P. Dufton for valuable comments. 
\end{acknowledgements}

\end{document}